\documentclass{article}
\begin{document}
\title{The status of the quantum dissipation-fluctuation relation and
Langevin equation}
\author{John C Taylor\\ Department of Applied Mathematics and Theoretical
 Physics,\\ Centre for Mathematical Sciences, Cambridge University,\\
Clarkson Road, Cambridge, UK}
\maketitle

\section*{Abstract}
I examine the arguments which have been given for
 quantum fluctuation-dissipation theorems. I distinguish between a
weak form of the theorem, which is true under rather general conditions,
and a strong form which requires a Langevin equation for its 
statement. I argue that the latter has not been reliably derived in
general.

\section{Introduction}
A quantum generalization of the Nyquist dissipation-fluctuation
relation was discussed by a number of authors many years ago.
 I shall refer explicitly
to the papers of Callen and Welton \cite{CW}  and of Senitzky
\cite{S}, and to the books
by Landau and Lifschitz \cite{LL}, Feynman and Hibbs \cite{FH}, Gardiner and Zoller
\cite{GZ}, Kogan \cite{K}, Balescu \cite{B} and Kreutzer \cite{Kr} .
I will distinguish between the weaker version (not including a Langevin
equation) of the relation first discussed by 
Callen and Welton and the stronger version
first formulated by  Senitzky, and examine the range of validity in each case. 
The latter author seeks to establish a quantum Langevin equation
containing 
a dissipation term and a noise term, connected by  the fluctuation-dissipation
 relation. The text books \cite{LL}, \cite{K}, \cite{B}, \cite{Kr}
follow \cite{CW}. The monograph \cite{GZ} includes a version of
\cite{S}. I also briefly review the formalism of
 \cite{FH}.

Another approach, which should contain much of the same information
is based on the master equation for the density operator (see \cite{GZ}).
I discuss this approach in section 4.

A {\it simple quantum system} 
 is envisaged in interaction with a {\it bath}
(the ``loss mechanism'') which has many degrees of freedom,
closely spaced energy levels, and is initially
in thermal equilibrium at temperature $T$.
The bath dissipates energy from the simple system
and at the same time feeds noise into it. The generalized Nyquist
relation connects the noise with the dissipation.

There is a special case in which all the equations of motion
for the bath degrees of freedom
are linear (that is,  the loss mechanism
consist of oscillators which are linearly coupled to the simple system).
  In this case, the theorem is certainly
well understood \cite{GZ}. This special case covers many important
applications, such as when the bath consists of photons
or phonons (if the interaction with the system is limited
to emission and absorption). But a bath consisting
of fermionic or spin degrees of freedom would not fall within
the special case.

The epithet ``quantum" implies that terms of order $\hbar$ are retained
in the expectation value of two noise operators. For consistency,
this implies that the noise term is indeed a non-commuting operator.
Since the Langevin equation gives the behaviour of the simple
system in terms of the noise, the simple system must also be
quantum. It has been stressed  in \cite{Gavish} that the expectation values
of the noise operators in their two different orderings have different
physical interpretations; so  it is impossible to ignore
the fact that they do not commute.

The quantum Langevin equation has been quoted in connection
with experiments \cite{Bloch} \cite{Beck} (see also \cite{K}, \cite{Lev})
 in which a Josephson junction is shunted
by a resistor. The Josephson phase difference $\delta$ is modeled
as the simple system, and the resistor is modeled  as the bath.
 The resistor $R$
tends to dissipate the Josephson junction voltage
 $U={\hbar \over 2e}\dot{\delta}$ because of its
conductance $1/R$. The noise current which couples to $U$ must be bilinear
in the electron creation and annihilation operators. Therefore the
equations of motion for  these electron operators are not linear
in the complete set of degrees of freedom ($U$ together with
the electron operators). So it is not obvious that the Langevin
equation can be justified.

 \section{The method of Callen and Weston}
In this section, I review a weak form of the fluctuation-dissipation
theorem, which is proved (originally in \cite{CW}) under rather general conditions, provided only
that departures from equilibrium are small.
 
The complete system 
 has quantum variables $q_i, p_i$ with
$i=1,...,N$ with $N$ large. If it is driven by an external
classical driving force $F(t)$,  the Hamiltonian is
$$H(q_i,p_i)-F(t)Q(q_i,p_i), \eqno(1)$$
where $Q$ is some particular combination of the dynamical variables.
(In one example, $F$ is an applied voltage and $\dot{Q}$ the
resulting current.)
The whole system is, in the absence of $F(t)$, in thermal equilibrium
at temperature $T$. The eigenvalues of $H$
are $E_n$, and are assumed to be closely spaced (so sums over $n$
can be approximated by integrals). 
 Thermal expectation values
are given by the distribution function
$$ \rho(H) = {\exp (-H/kT) \over \hbox{tr} \exp(-H/kT)}. \eqno(2)$$

Following Landau and Lifshitz \cite{LL}, we expect,
 provided that $F$ is small enough,
a linear response of the form
$$\langle \dot{Q}_F(t) \rangle \equiv \hbox{tr}(\rho \dot{Q}_F)=\int_0^{\infty}dt'\chi(t-t')F(t'), \eqno(3)$$
(the suffix $F$ on $Q_F$ indicates that this is the response to
the external force $F$; I shall use $Q$ with no suffix to denote
a fluctuation in the absence of $F$). Equation (3)   defines
 a ``linear response
function'' or
 ``generalized susceptibility'' $\chi$ (the inverse of the
impedance). Note that it is the expectation value of $Q_F$ which appears in
(3), so the whole equation refers to classical quantities.

For the case where
$$ F(t)=\Re [F_0 \exp (i\omega t)], \eqno(4)$$
we have (in terms of Fourier transformed function, denoted by $\tilde{~}$
$$i\omega \langle {\tilde{Q}}_F(\omega) \rangle=\tilde{\chi}(\omega)F_0, \eqno(5)$$
where
$$\tilde{\chi}(-\omega)=-\tilde{\chi}^*(\omega). \eqno(6)$$
The power provided by $F$ and dissipated into the system is
$$W=\langle \partial H/\partial t\rangle=-\langle Q_F \rangle \dot{F}(t) \eqno(7)$$
and its mean value (over time) is
$$\overline{W} ={1\over 2}\Re\{\tilde{\chi}(\omega)\}|F_0|^2. \eqno(8)$$

But we can also calculate $\overline{W}$ by applying lowest order perturbation theory
to the perturbing Hamiltonian $-QF$. The result can be expressed in terms
of the expectation value
$$S(\omega,T)\equiv {1\over 2\pi}\int dt \exp(-i\omega t)\hbox{tr}\{\rho Q(0)Q(t)\} \eqno(9)$$
Comparison of this result with (8), 
 yields the theorem
$$ S(\omega,T)={2\over  \pi}\left[ {\hbar \omega \over
\exp(\hbar\omega/kT)-1}\right]\Im \{\tilde{\chi}(\omega)\}. \eqno(10)$$

If we choose instead in (9) the symmetrised product of operators, we get
$$S_{sym}\equiv {1\over 2} [S(\omega,T)+S(-\omega,T)]={1\over \pi}N(\omega,T)
\Re \{ \tilde{\chi}(\omega)\}, \eqno(11)$$
where
$$ N(\omega)=\hbar \omega \left[{1\over \exp(\hbar \omega /kT)-1}+{1\over 2}
 \right] \eqno(12)$$
is the Bose distribution function.

I call (10) or (11) the weak fluctuation-dissipation theorem.
It is very general and simple, but it has the a
limitation, which I shall now explain.
If $Q$ were a classical quantity, one could have, more generally than (3),
$$\dot{Q}_F=\int_0^{\infty}dt'\chi(t-t')F(t'). \eqno(13)$$
Then it would be natural to define a fluctuation $f$ in the force related to
the fluctuation $Q$ by
$$\dot{Q}(t)=\int_0^{\infty}dt'\chi(t-t')f(t'). \eqno(14)$$
It is then possible to deduce from (10) and (14) a formula for 
$$ S_f \equiv {1\over 2\pi} \int dt \exp(-i\omega t)\hbox{tr} \{ f(0)f(t)\}. \eqno(15)$$
 But in the quantum case,
using only (3), one cannot take these steps, and so cannot go beyond the
fluctuation-dissipation relation (10).

 Some authors (for example
 Callen and Welton \cite{CW},  Landau and Lifshitz \cite{LL},
and Kogan \cite{K}) tacitly assume (14) even in the quantum case.
Landau and Lifshitz write ``It is convenient to write the formula [(14)]
{\it as if} $Q$ were a classical quantity'' (my italics).
These authors also call $f$ a ``fictitious random force''.
But, in my opinion, $f$ is not fictitious; it is a real dynamical
force exerted on the degree of freedom $Q$ by the other degrees of freedom in
$q_i,p_i$. In order to exhibit this, and to justify (14) in the quantum case
(if that is possible), one needs a Langevin equation, and this is
the subject of the next section.

\section{Senitzky's derivation of a Langevin equation}

In this method, one degree of freedom $Q,P$ is singled out,
representing a {\it simple system} 
 and this interacts weakly with a {\it  bath}
with degrees of freedom $q_i, p_i$, $i=1,...,N$. 
It would be possible for the $q_i,p_i$ variables to be quantum,
while $Q,P$ was a classical degree of freedom; but I am concerned
with the situation in which $Q,P$ are quantum as well.
For quantum effects in the simple system to be important, it
presumably must be microscopic (for instance a single atom), perhaps
mesoscopic (like a very small Brownian particle), or perhaps the phase
of a superconducting condensate (\cite{Bloch} \cite{Beck}).

For simplicity taking the simple system to be an oscillator, the Hamiltonian
is
$$ H={1\over 2m}P^2+{1\over 2}m\Omega^2 Q^2+H_{B}(q_i,p_i)+H_{int}
(Q,P;q_i,p_i)-QF(t),
\eqno(16)$$
with the interaction
$$H_{int}=-\alpha Q K(q_i,p_i). \eqno(17)$$
Here $F$ is a classical driving force, $\alpha$ is a small parameter,
 and $K$ is some function of the
bath variables. The suffix $B$ stands for ``bath".
 The choice $Q$ rather than $P$ in (16)
is somewhat arbitrary, but it is important that this term is linear in $Q,P$.
Some parts of the argument would go through for a general potential
$V(Q)$ instead of $m\Omega^2Q^2/2$.
The Hamiltonian $H$ is positive provided that
$$ 2m\Omega^2H_{B} \geq \alpha^2 K^2. $$

Note that I have used the same notation as in section 2, but it
refers to a slightly different physical model.

The Heisenberg picture is used. The equations of motion are
$$\dot{P}=-m\Omega^2 Q+\alpha K +F,~~~~\dot{Q}={1\over m}P.
 \eqno(18)$$
 
The special case in which $H_{B}$ is quadratic in its variables and
$K$ is linear (so that the whole assembly is made up of harmonic
oscillators and all the equations of motion are linear)
 is simple and well-studied (see for instance \cite{GZ}),
 and there is
 no doubt about the validity of the fluctuation-dissipation
 theorem. But it is not clear that this model
covers all physical examples. The argument of \cite{S} however
claims to be more general. 

It might be argued  (see for example \cite{FH}) that,
 close to equilibrium, we can expand
potentials about their equilibrium values and a quadratic approximation
for $H$ is necessarily justified. But if $K$ is itself a quadratic function,
then this approximation would decouple the simple system from the bath
altogether. For example, if any of the bath variables are fermionic,
then $K$ cannot be a linear function of them. 

We define operators with a super-fix $(0)$ to satisfy the equations of motion
when $\alpha=0$, and to coincide with the true operators at some
initial time $t_0$; so for example
$$
K^{(0)}(t)= K(t;\alpha=0),~~~K^{(0)}(t=t_0)=K(t=t_0).
$$
$$\dot{P}^{(0)}=-m\Omega^2 Q^{(0)},~~m\dot{Q}^{(0)}=P^{(0)};~~~
P^{(0)}(t_0)=P(t_0),~~Q^{(0)}(t_0)=Q(t_0). \eqno(19)$$

By using the Heisenberg equations of motion, Senitzky (equation (16) of
 \cite{S}) obtains
$$K(t)=K^{(0)}(t)$$ $$-{\alpha \over  \hbar^2}\int_{t_0}^t dt' \int_{t_0}^{t'} dt''
U(t'-t)[K(t'),[K(t''),H_{B}(t'')]Q(t'')]U(t-t'), \eqno(20)$$
where
$$ U(t'-t)=\exp\{-i(t'-t) H^{(0)}/\hbar\}. \eqno(21)$$
Neglecting terms of higher order in $\alpha$, we replace $K$ by $K^{(0)}$
and $H_B$ by $H^{(0)}_B$ on the right hand side of (20).
We may also neglect the commutators of $Q$ with $H^{(0)}_B$ and with
$K^{(0)}$ since these are relatively  $O(\alpha)$. Using the time translation
operator $U$ (I differ from \cite{S} slightly at this point), we finally get
$$K(t) \approx K^{(0)}(t)-i\alpha\int_{t_0}^t dt''C(t,t'')Q(t''), \eqno(22)$$
where (with $\tau=t+t''-t'$)
$$C(t,t'')\equiv {1 \over \hbar}\int_{t''}^t d\tau [K^{(0)}(t),
\dot{K}^{(0)}(\tau)]=-{1 \over \hbar}[K^{0}(t),K^{0}(t'')].
 \eqno(23)$$

Equation (22) may now be inserted into (18) to obtain the approximate
linear equations for $Q$ and $P$:
$$\dot{P}(t)=-m\Omega^2Q(t)+f(t)+\alpha K^{0}(t)-i\alpha^2 
\int_{t_0}^t dt''C(t,t'')Q(t''),$$
$$ \dot{Q}(t)={1\over m}P(t). \eqno(24)$$
Here the $K^{0}$ term may be interpreted as quantum noise,
and the $C$ term contains a dissipative part.

At this stage, $C$ in (23) is a quantum operator, not a $c$-number,
and it is in general a function of the two variables $t,t''$ not just their
difference. In the special case when $H_{B}$ is quadratic and $K$ is linear
(the oscillator case), $C$ is a $c$-number and is a function only of
$t-t''$, and is independent of $T$. But the question is whether
(24) can be simplified for a general $H_B$ and $K$.

At this point, Senitzky \cite{S} makes the approximation of replacing $C$ by its expectation value:
$$C(t-t'')\approx \hbox{tr}\{ \rho  C(t,t'')\}\equiv c(t-t'',T) , \eqno(25)$$
where now
$$ \rho = {\exp(-H_B^{(0)}/kT) \over \hbox{tr} \exp(-H_B^{(0)}/kT)}. \eqno(26)$$
Notice that this $\rho$ is not quite the same as that defined in the 
previous section in (2).

In justifications for this approximation, Senitzky \cite{S} writes
 ``we ignore the quantum-mechanical properties of the
loss-mechanism [that is, the bath] 
in terms of higher order than the second", and ``all our final results
(but nor necessarily the intermediate steps) will be expectation values 
with respect to the loss-mechanism; and since the term affected involves
only second and higher order interactions, only the higher order
quantum mechanical effects are neglected in the final result".
I have not understood the distinction between higher order
effects (higher order in $\alpha$) and higher order quantum mechanical
effects. It does not seem to me that the terms neglected in the approximation
(24) are higher order in $\alpha$ or in $\hbar$ than the terms retained.

{\it If} we make the approximation (25) and insert it into (24),
 we obtain an equation of Langevin type.
If we choose $t_0=-\infty$ and take Fourier transforms (denoted by
$\tilde{Q}$ etc), we obtain
$$[m(\Omega^2-\omega^2)+G (\omega,T)]\tilde{Q}(\omega)
\equiv i\omega Z(\omega,T)\tilde{Q}(\omega)
=\alpha\tilde{K}(\omega)+\tilde{F}(\omega), \eqno(27)$$
where
$$G(\omega,T)=\alpha^2 \int d\omega'{1\over \omega'-\omega-i\epsilon}~
\tilde{c}(\omega',T),
\eqno(28)$$
where $\tilde{c}$ is the Fourier transform of $c$ defined in (25).
So, in this approximation, there does exist a $c$-number impedance $Z$
and $\tilde{Z}$ is the inverse of the susceptibility $\tilde{\chi}$
defined in section 2 in equation (5). The real part of $\tilde{Z}$
is the resistance giving the dissipation.
But, without the approximation
(25), the $C$ term in the equation of motion is a quantum operator.

The term $\alpha \tilde{K}$ in (27) is interpreted as noise and 
plays the role of the force fluctuation $f$ hypothesized at the end of
section 2.
The noise spectrum is (choosing the symmetrized product)
$$S^{(f)}_{sym}(\omega,T)= {\alpha^2 \over 4\pi}\int dt \exp(-i\omega t)\hbox{tr}\{\rho (
{K}(0){K}(t)+{K}(t){K}(0)\}
,\eqno(29)$$
where the superfix $(f)$ denotes the noise in $f$ as opposed to $Q$.
Then the fluctuation-dissipation theorem ({\it if} the approximation (26) were valid)
would be
$$ S^{(f)}_{sym}(\omega,T)={1\over \pi}[\Im Z(\omega,T)] N(T,\omega), \eqno(30)$$  where $N$ is the Bose function defined in (12).
The theorem could alternatively be expressed in terms of the fluctuation
$ Q$ produced by the noise $K$, by using
$$i\omega Z(\omega)\tilde{Q}=\alpha \tilde{K}(\omega), $$
where $Z$ is defined in (27).

\section{The argument of Feynman and Hibbs}

A fluctuation-dissipation theorem is derived in section 12.9 of \cite{FH}
(see also \cite{Sc}). The argument is very similar that of Senitsky
except that it uses Feynman's path integrals instead of Heisenberg
equations of motion. It assumes that the Lagrangian is
of second degree  (so that the bath degrees of freedom can
be explicitly integrated out). Feynman and Hibbs appear to argue
that such a second degree Lagrangian is very general.

\section{The master equation}

A different approach to the same physical problem is via the
master equation for the density operator $\rho$ (see for example
\cite{GZ} chapter 5).  For a critique of this method in the quantum case
see \cite{vanK}.

Using now the interaction picture
(instead of the Heisenberg picture), the density operator
obeys the equation
$$ \hbar \dot{\rho}(t)=-i[H_{int}(t),\rho(t)], \eqno(31)$$
where $H_{int}$ is defined in (17).
Iterating this equation,
$$ \dot{\rho}(t)=-{i\over \hbar}[H_{int}(t),\rho(t_0)]-
{1\over \hbar^2}\int_{t_0}^t
dt'[H_{int}(t),[H_{int}(t'),\rho(t')]]. \eqno(32)$$
We assume the initial condition that the density operator factorizes
into a bath part and a system part
$$\rho(t_0)=\rho_{B}\otimes \rho_{S}. \eqno(33)$$
We would normally choose $\rho_{B}$ to be the Boltzmann distribution.

From (32) we can deduce equations for expectation values of operators.
For example
$$ {d \over dt}\hbox{tr}\{\rho(t)P(t)\}=- m\Omega^2
\hbox{tr}\{\rho(t)Q(t)\} +\alpha\hbox{tr}\{\rho(t_0)K(T)\}\dot{P}(t)$$
$$+{i\alpha^2\over \hbar}\int_{t_0}^t dt'\hbox{tr}\{\rho(t')[K(t),K(t')]Q(t')\},
\eqno(34)$$
$$ {d \over dt}\hbox{tr}\{\rho(t)Q(t)\}={1\over m}
\hbox{tr}\{\rho(t)P(t)\}. \eqno(35)$$

{\it{If}} the loss mechanism consists of a set of oscillators, and if
$K$ is linear in these oscillator operators, then the commutator
$[K(t),K(t')]$ in (34) is a $c$-number.
In that case, the last term on the right in (34) contributes a
dissipative term to the equations of motion for the expectation values
of $Q$ and $P$, consistently with the expectation value of the 
Langevin equation (24) in section 3.

In many cases (for instance for the oscillator model),
 the expectation value of the noise term (the second term on the right)
in (34)  vanishes.
But we may look for the effect of noise by taking
expectation values of products of two operators. 
For example, we could obtain a differential equation like (34) for
$$ \hbox{tr}\{\rho(t)P(t)P(t)\}, \eqno(36)$$
and this would include expectation values of terms bilinear in $K$
which would not vanish. However, from solutions of the 
Langevin equations (24), one could get (in the Heisenberg picture)
expectation values like
$$\hbox{tr}\{\rho_H P_H (t_1)P_H (t_2)\} \eqno(37)$$
 (where the suffix  $H$ emphasizes the Heisenberg picture).
Such unequal time correlators seem to appear less naturally
in the master equation formalism (see for example section 5.2.1
of \cite{GZ}).

The question at issue in this paper is whether one can obtain
an approximate Langevin-like equation  when the commutator $[K(t),K(t')]$
in (34) is {\it not} a $c$-number. This would be possible if the
density matrix factorized (approximately) at all times:
$$ \rho(t) \approx \rho_S (t) \otimes \rho_{B}(t), \eqno(38)$$
where the suffices refer to {\it simple system} and {\it bath} respectively.
To investigate the validity of this approximation, we can insert
(38) into (31) and trace over the bath variables, to get
(with the Hamiltonian (16))
$$ \hbar \dot{\rho}_S(t)\approx -i \hbox{tr}_{B}(\rho_{B}K)[Q,\rho_S].
\eqno(39)$$
In many cases, in particular in the simple case when the bath
is a set of oscillators and $K$ is linear in these variables (which
is the simplest possible case, and for which the method of section 3
succeeds)
the trace on the right hand of (39) is is zero. This means that
 $\rho_S(t)\approx\rho_S(t_0)$.
If it were legitimate to insert this approximation into the last term in
(32), we would get no integral equation for $\rho_B(t)$ at all. Thus the approximation
(38) seems to be unreasonable.

\section {Comments}
 As mentioned in the Introduction, the quantum fluctuation-dissipation
 relation has been referred to
in connection with experiments which might be thought to be sensitive to
vacuum energy \cite{Bloch} \cite{Beck}.
I have argued in this note that one should not appeal
 the quantum fluctuation-dissipation relation without checking
if it is valid for the case in point.

What is the interpretation of  of the Bose distribution 
function (12) appearing in the fluctuation-dissipation relations? First of all,
it is present only if the symmetrized product (11) is used (see
\cite{Gavish}). In the ideal case where the equations of motions
are linear, and so the commutator $C$ in (23) is a $c$-number and the
dissipation is independent of $T$, the Bose function (12) gives
the complete temperature dependence of the noise spectrum (30).
In this case, there is little doubt that (10) corresponds to the
oscillators in the bath. If these are three dimensional, we would expect
the dissipation in $G$ in (27) to be proportional to $\omega^2$.

Beyond this special case, the situation is less clear. I have argued above that
the Langevin equation has not been proved in general. But even if we
{\it assume} that the approximation (25) is justified, the factor $N$ in (12)
does not give the $T$-dependence of the noise spectrum, because the dissipation
itself will be in general $T$-dependent. {\it If} the fluctuation-dissipation
relation (11) had been justified generally, it would cover cases where
the dissipation and noise were due to a bath of fermionic systems, and then the
 Bose factor $N$ in (12) could not possibly represent the physics of the bath.
\section*{Acknowledgments}
I am grateful to Christian Beck, Roger Koch,  Crispin Gardiner, Fred Green
and the referees
for their extensive advice, patiently given. But none of these
is responsible for the shortcomings of this paper.

\end{document}